\documentclass[a4paper,12pt]{article}

\usepackage{amsmath}\usepackage{amssymb}
\usepackage{latexsym}\usepackage{cite}

\usepackage{amsmath}
\usepackage[dvips]{graphicx}
\usepackage{pstricks}
\usepackage{color}
\usepackage{bm}

\usepackage{graphicx}

\definecolor{Red}{rgb}{1.00, 0.00, 0.00}
\definecolor{Green}{rgb}{0.00, 1.00, 0.00}
\definecolor{Blue}{rgb}{0.00, 0.00, 1.00}
\definecolor{Cyan}{rgb}{0.00, 1.00, 1.00}
\definecolor{Mymagenta}{rgb}{0.3, 0.00, 1.00}%
\definecolor{Magenta}{rgb}{1.00, 0.00, 1.00}
\definecolor{DeepSkyBlue}{rgb}{0.00, 0.75, 1.00}
\definecolor{DarkGreen}{rgb}{0.00, 0.39, 0.00}
\definecolor{SpringGreen}{rgb}{0.00, 1.00, 0.50}
\definecolor{Mygreen}{rgb}{0.00, 0.72, 0.00}
\definecolor{DarkOrange}{rgb}{1.00, 0.55, 0.00}
\definecolor{OrangeRed}{rgb}{1.00, 0.27, 0.00}
\definecolor{DeepPink}{rgb}{1.00, 0.08, 0.57}
\definecolor{DarkViolet}{rgb}{0.58, 0.00, 0.82}
\definecolor{SaddleBrown}{rgb}{0.57, 0.27, 0.07}
\definecolor{Black}{rgb}{1.00, 1.00, 1.00}
\definecolor{Ablue}{rgb}{0.10, 0.1, 1.00}

\newcommand{\be}{\begin{equation}}
\newcommand{\ee}{\end{equation}}

\newcommand{\ka}{\kappa}

\def\beq{\begin{equation}}
\def\eeq{\end{equation}}
\def\beqr{\begin{eqnarray}}
\def\eeqr{\end{eqnarray}}
\def\pl{\partial}

\def\al{\alpha}

\def\Ga{\Gamma}
\def\ga{\gamma}
\def\de{\delta}

\def\ka{\kappa}
\def\si{\sigma}
\def\Si{\Sigma}
\def\te{\theta}
\def\Te{\Theta}
\def\La{\Lambda}
\def\lam{\lambda}

\def\om{\omega}
\def\ep{\epsilon}

\def\sq{\sqrt}

\def\l{\left (}
\def\r{\right )}

\def\fr{\frac}
\def\la{\label}
\def\hs{\hspace}
\def\vs{\vspace}

\def\ran{\rangle}
\def\lan{\langle}
\def\ov{\overline}
\def\tl{\tilde}
\def\tm{\times}

\DeclareMathOperator{\e}{e}

\begin{document}

\begin{flushright}
January 15, 2015 \\
\end{flushright}

\vs{1.5cm}

\begin{center}
{\Large\bf

  Natural Inflation  from 5D SUGRA \\

 \vs{0.2cm}

 and Low Reheat Temperature}

\end{center}

\vspace{0.5cm}
\begin{center}
{\large
{}~Filipe Paccetti Correia$^{\hs{0.5mm}a}$\footnote{E-mail: fcorreia@deloitte.pt},{}~
Michael G. Schmidt$^{\hs{0.5mm}b}$\footnote{E-mail: m.g.schmidt@thphys.uni-heidelberg.de}, and
{}~Zurab Tavartkiladze$^{\hs{0.5mm}c}$\footnote{E-mail: zurab.tavartkiladze@gmail.com}
}
\vspace{0.5cm}

$^a${\em Deloitte Consultores, S.A., Pra\c{c}a Duque de Saldanha, 1 - 6$\hs{0.4mm}^o$\!\!, 1050-094
Lisboa, Portugal}$^{\hs{0.5mm}}$\footnote{Disclaimer: This address is used by F.P.C. only for the purpose of indicating his
professional affiliation. The contents of the paper are limited to Physics and in no ways represent views
of Deloitte Consultores, S.A.}

$^b${\em Institut f\"ur Theoretische Physik,
Universit\"at Heidelberg,
Philosophenweg 16, \\ 69120 Heidelberg, Germany }

$^c${\em Center for Elementary Particle Physics, ITP, Ilia State University, 0162 Tbilisi, Georgia}
\end{center}

\vspace{0.6cm}

\begin{abstract}


Motivated by recent cosmological observations of a possibly unsuppressed primordial tensor component $r$ of inflationary perturbations,
we reanalyse in detail the 5D conformal SUGRA originated natural inflation model of Ref. \cite{Paccetti:2005zm}.
The model is a supersymmetric variant of 5D extra natural inflation, also based on a shift symmetry, and leads to the
potential of natural inflation.
Coupling the bulk fields generating the inflaton potential via a gauge coupling to the inflaton with
brane SM states we necessarily obtain a very slow gauge
inflaton decay rate and a very low reheating temperature $T_r\stackrel{<}{_\sim }{\cal O}(100)$~GeV.
Analysis of the required number of e-foldings (from the CMB observations) leads to values of $n_s$
in the lower range of present Planck 2015 results.
Some related theoretical issues of the construction, along with phenomenological
and cosmological implications, are also discussed.

\end{abstract}

\newpage

\section{Introduction}

Inflation solves the problems of early cosmology in a natural way \cite{first-infl} and besides
that produces a primordial fluctuation spectrum \cite{Mukhanov:1981xt}
 which allows to discuss
structure formation successfully. In detailed models (i) a sufficient number of
$\e$-folds for the inflationary phase has to be produced, (ii) guided by bounds presented recently by the
Planck Collaboration \cite{Planck:2015xua},  the
cosmic background radiation and  a spectral index  $n_s=0.968\pm 0.006$  should be generated.\footnote{In the original version of
this paper (see v1 of arXiv:1501.03520) we had the 2013 value \cite{Ade:2013zuv} $n_s=0.9603\pm 0.0073$ which being about a standard
deviation below this value makes quite a difference for our analysis.}
And (iii), the normalization of fluctuations has to be
reproduced. Rather flat potentials for the inflaton field lead to the ``slow roll''
needed for (i). Such potentials appear naturally in (tree level) global
supersymmetric models; higher loop corrections can be controlled, but the
inclusion of supergravity easily produces an inflaton mass of the order of the
Hubble scale.

In models with an extra dimension the fifth component of a $\text{U}(1)$ gauge
field entering in a Wilson loop operator can act as an inflaton field of pseudo
Nambu-Goldstone type which is protected against gravity corrections and avoids a
transplanckian scale \cite{ArkaniHamed:2003wu}, \cite{Kaplan:2003aj}, present in the
original model of ``natural inflation"  \cite{Freese:2014nla}.
 We have presented such a model \cite{Paccetti:2005zm}
based on 5D conformal SUGRA on an orbifold $\text{S}^1\!/\mathbb{Z}_2$ with a predecessor based on global
supersymmetry with a chiral ``radion'' multiplet on a circle in the fifth dimension~ \cite{Hofmann:2003ag}.
We also made the interesting observation that
a spectral index $n_{\text{s}} \sim 0.96$ as observed recently [different from a value very close to one usually obtained in
 straightforward SUSY hybrid inflation \cite{Okada:2014lxa}], is obtained rather generically in gauge inflation. Actually, in the
supersymmetric formulation we have a complex scalar field which besides the
gauge inflaton $A_5^1$ contains a further ``modulus'' field $M^1$ which also might
allow for successful inflation \cite{Paccetti:2005zm}. The main difference between the two inflation types is that gauge
inflation leads to a large tensor to scalar ratio $r(\sim 0.12$ in \cite{Paccetti:2005zm})
whereas modulus inflation leads to very small $r(\stackrel{<}{_\sim} 10^{-4}$ in \cite{Paccetti:2005zm}).\footnote{Genuine two field inflation
was discussed in ref. \cite{2a}. The two basic inflation types depending on initial conditions turn out to be still like in \cite{Paccetti:2005zm}.
Since inclusion of the $M^1$ modulus into the inflation process is fully legitimate, one can reserve this scenario as an
alternative with a tiny tensor perturbations, if it should be.}
 Recently the BICEP2 data \cite{Ade:2014xna} gave strong indication of a large ratio  $r=0.2^{+0.07}_{-0.05}$
 though recent joint analysis of BICEP2/Keck and Planck \cite{Ade:2015tva} gave a reduced  upper bound
  $r\stackrel{<}{_\sim }0.12$,\footnote{Earlier, Planck's intermediate results \cite{Adam:2014bub}
 noted about a possible ordinary dust contribution instead of the light polarization effect really due to gravitational waves.} with
the likelihood curve for $r$ having a maximum for $r\simeq 0.05$.
Because of this,
 we here consider the gauge inflation of ref. \cite{Paccetti:2005zm} again with particular emphasis on the required length of inflation.
  The well known $62$  $e$-folds solving the horizon problem will turn out to require a substantial expansion during the reheating period within the natural inflation scenario emerged from 5D SUGRA.

Let us present the organization of the paper and summarize some of the results. In Sec. \ref{nat} we perform a
detailed analysis of natural inflation with
$\cos $-type potential. For the calculation of the spectral index $n_s$ and the tensor to scalar ratio $r$, we use a second order
approximation with respect to slow roll (SR) parameters. Since these quantities ($n_s, r$) are determined
at the point where the SR parameters are tiny, this approximation is sufficient for all practical purposes.
However, near  the end of the inflation, when SR breaks down, we perform an accurate numerical determination
of the point via the condition $\ep_H=1$ on the Hubble slow roll parameter (see \cite{Stewart:1993bc}-\cite{Liddle:1994dx} for definitions). This
is needed to compute, with desired precision, the number of $e$-foldings ($N_e^{\rm inf}$) before the end
of the inflation. We carry out our analysis by using recent fresh data \cite{Planck:2015xua} for $n_s$, $N_e$, the amplitude of curvature perturbations and
bounds for $r$ \cite{Ade:2015tva}. Our results are in agreement with a recent analysis of natural inflation by Freese and Kinney
(see $2^{\rm nd}$ citation in \cite{Freese:2014nla}).
For various cases we have also calculated the reheat temperature, which we use later on for contrasting with
 our 5D SUGRA emerged inflation scenario requiring a very low reheat temperature.

%
%

In Sec. \ref{NATinf5D} and Appendix \ref{KK}
we shortly review our model of ref. \cite{Paccetti:2005zm} in a more self-contained way
 and discuss how natural inflation emerges  from $5$D SUGRA.
Using a superfield formulation, we do not need to go into the details of the component expressions in
conformal 5D SUGRA of Fujita, Kugo and Ohashi (FKO)  \cite{Fujita:2001kv}. Indeed this emerged
from our discussion \cite{Paccetti:2004ri, ef4D-sugra} (see also Ref. \cite{Abe:2005ac})  bringing the
5D conformal SUGRA formulation closer to the 4D global SUSY language \cite{ef4D-sugra}.
We concentrate here on gauge inflation, i.e. on the case $M^1=0$ (stabilized moduli in the origin
or a choice of initial conditions\footnote{For a discussion
of moduli stabilization in the superfield formalism within 5D SUGRA see \cite{Correia:2006vf}. For a choice of initial conditions leading approximately
to $M^1=0$ see Ref. \cite{2a}.}).
In section \ref{NATinf5D}, discussing the realization of natural inflation within 5D SUGRA, we present a new mechanism for
inflaton decay,  which eventually leads to the reheat of the Universe.
Note that, besides a specific string theory realization \cite{Blumenhagen:2014gta}, the inflaton decay and reheating has never been
discussed before in the context of natural inflation.
 We show that the inflaton's slow decay is a natural consequence of the 5D
construction (with consistent UV completion),
 being realized by couplings of the heavy bulk
supermultiplets generating the inflaton potential through their gauge coupling  with brane SM states. Since the inflaton
decay proceeds by 4-body decay and the decay width is strongly
suppressed by the 2-nd power of the tiny $U(1)$ gauge coupling constant\footnote{\la{wgc}From a very recent paper \cite{delaFuente:2014aca}
we learned that the `weak gravity conjecture'  (going back to Ref. \cite{ArkaniHamed:2006dz})
based on magnetically charged black hole considerations and the dangerous neighborhood to a global symmetry,
applies in disfavor of gauge (extranatural) inflation and might explain difficulties to embed the model in string theory.}
 (of the gauge inflaton-charged fields) and a relatively small inflaton mass
coupled to the intermediate bulk fields,
a strong suppression of the reheat temperature $T_r$ comes out naturally. Our 5D SUGRA
construction allows us to make an estimate $T_r\sim 0.34\rho_{reh}^{1/4}\sim |\lam |^2\tm 100$~GeV
(where $\lam \stackrel{<}{_\sim }1$ is a brane Yukawa coupling).  At the end of  Sec. \ref{NATinf5D}
we show that, by the parameters we are dealing with, preheating is excluded within the considered scenario.

Appendix \ref{KK} discusses the Kaluza-Klein spectrum of the fields involved, as well as the SUSY breaking effects for  brane fields.
We also perform a derivation of higher dimensional operators involving the inflaton $\phi_{\Te }$ and light (MSSM)  states relevant
for the inflaton decay. As it turns out, the dominant decay channel is  $\phi_{\Te }\to llhh$ (with $l$ and $h$ denoting SM lepton and Higgs
doublets respectively).
Sec. \ref{conc} includes a discussion and concluding remarks about some related issues.

\section{Natural inflation}
\la{nat}

In this section we analyse  inflation with the potential of natural inflation \cite{Freese:2014nla} given by:
\beq
{\cal V}={\cal V}_0\l 1+\cos(\al \phi_{\Te})\r ~,
\la{pot}
\eeq
where $\phi_{\Te}$ is a canonically normalized real scalar field of inflation. In the concrete scenario of
Ref. \cite{Paccetti:2005zm}, we focus later on, the inflaton originates from a $5$D gauge superfield, while the parameters/variables of
(\ref{pot}) are derived through the underlying 5D SUGRA. See Eqs. (\ref{pot-5D-params}), (\ref{rel-infl-A5}), (\ref{Vinfl})
and also the comment underneath Eq. (\ref{Vinfl}).

The slow roll parameters ("VSR" - derived through the inflaton potential) are given by
$$
\ep =\fr{M_{Pl}^2}{2}\l \fr{{\cal V}'}{{\cal V}}\r^2=\fr{(M_{Pl}\al)^2}{2}\tan^2\fr{\al \phi_{\Te}}{2}
$$
$$
\eta =M_{Pl}^2\fr{{\cal V}''}{{\cal V}}=\fr{(M_{Pl}\al)^2}{2}\l \tan^2\fr{\al \phi_{\Te}}{2}-1\r =\ep-\fr{1}{2}(M_{Pl}\al)^2~,
$$
\beq
\xi =M_{Pl}^4\fr{{\cal V}'{\cal V}'''}{{\cal V}^2}=-(M_{Pl}\al)^4\tan^2\fr{\al \phi_{\Te}}{2}=-2(M_{Pl}\al)^2\ep ~,
\la{ep-eta}
\eeq
where $M_{Pl}=2.4\cdot 10^{18}$~GeV is the reduced Planck mass.
In order to make notations compact, for the VSR parameters we do not use the
subscript `V' (denoting them by $\ep, \eta , \xi ,\cdots $). However, for HSR parameters (derived through the Hubble parameter)
 we use subscript `H' (e.g. $\ep_H, \eta_H , \xi_H ,\cdots $), as adopted in
 literature \cite{Stewart:1993bc}, \cite{Liddle:1994dx}, \cite{Peiris:2003ff}.
 The number of e-foldings during inflation, i.e. during exponential expansion,
denoted further by $N_e^{\rm inf}$, is calculated as

\beq
N_e^{\rm inf}=\fr{1}{\sqrt{2}M_{Pl}}\int_{\phi_{\Te}^e}^{\phi_{\Te}^i} \fr{1}{\sqrt{\ep_H}} d\phi_{\Te }~.
\la{exact-Ninf}
\eeq
In this exact expression the HSR parameter $\ep_H$ (defined below), participates.
The point $\phi_{\Te}^e$, at which inflation ends, is determined by the condition $\ep_H=1$.
The point $\phi_{\Te}^i$ corresponds to the begin of the inflation. Also further, symbols with superscript or subscript 'i' will correspond to values at the beginning of the inflation, while
superscript/subscript 'e' will indicate end of the inflation.

The observables   $n_s$ and $r$ depend on the value of $\phi_{\Te}^i$  (the point at which scales cross the horizon).
This allows to determine $\phi_{\Te}^i$ as follows.
Via HSR parameters, the expressions for $n_s$ and $r$ are given by \cite{Stewart:1993bc}, \cite{Liddle:1994dx}, \cite{Peiris:2003ff}:
$$
n_s=1-4\ep_{Hi}+2\eta_{Hi}-2(1+C)\ep_{Hi}^2-\fr{1}{2}(3-5C)\ep_{Hi}\eta_{Hi}+\fr{1}{2}(3-C)\xi_{Hi} ,
$$
\beq
r=16\ep_{Hi}\l 1+2C(\ep_{Hi}-\eta_{Hi})\r ~,~~~~{\rm with} ~~C=4(\ln 2+\ga )-5\simeq 0.0815~,
\la{ns-r-HSR}
\eeq
where we have limited ourself with second order corrections.
The HSR parameters $\ep_H, \eta_H, \xi_H$ are given by:
\beq
\ep_H=2M_{Pl}^2\!\l \!\fr{H'}{H}\!\r^2 \!,~~~~~\eta_H=2M_{Pl}^2\! \fr{H''}{H} ,~~~~~\xi_H=4M_{Pl}^4\!\fr{H'H'''}{H^2}~,
\la{def-HSR-par}
\eeq
with the Hubble parameter $H$ and it's derivative with respect to the inflaton field.
The subscript $'i'$ in (\ref{ns-r-HSR}) indicates that the parameter is defined at the point at which scales cross the horizon.
As it turns out, at this scale the slow roll parameters are small and second order corrections in $n_s$ and $r$ are small and the approximations
made in (\ref{ns-r-HSR}) are pretty accurate.
 Exact  relations between VSR ($\ep , \eta, \xi , \cdots $) and HSR parameters
 ($\ep_{H}, \eta_{H}, \xi_{H}, \cdots $) are given by \cite{Stewart:1993bc}, \cite{Liddle:1994dx}, \cite{Peiris:2003ff}:
 $$
 \ep=\ep_H\l \fr{3-\eta_H}{3-\ep_H}\r^2 ,~~~~\eta=\fr{3(\ep_H +\eta_H)-\eta_H^2-\xi_H}{3-\ep_H} ,
 $$
 \beq
 \xi =3\fr{3-\eta_H}{(3-\ep_H)^2}\l 3\ep_H\eta_H+\xi_H(1-\eta_H)-\fr{1}{6}\si_H\r ,~~~~{\rm with}~~~\si_H=4M_{Pl}^4\ep_H\fr{H^{\rm (iv)}}{H}.
 \la{exact-rel-HSR-VSR}
 \eeq
When the slow roll parameters are small, from (\ref{exact-rel-HSR-VSR}), the HSR parameters to a good approximation can be expressed in terms of VSR
parameters as
 $$
 \ep_H\simeq \ep -\fr{4}{3}\ep^2+\fr{2}{3}\ep \eta ~,~~~\eta_H\simeq \eta -\ep +\fr{8}{3}\ep^2-\fr{8}{3}\ep \eta +\fr{1}{3}\eta^2+\fr{1}{3}\xi ,
 $$
 \beq
 \xi_H\simeq 3\ep^2-3\ep\eta+\xi .
 \la{approx-rel-HSR-VSR}
 \eeq
Using these approximations  in (\ref{ns-r-HSR}), we can write  $n_s$ and $r$ in terms of VSR parameters:
$$
n_s=1-6\ep_i+2\eta_i +\fr{2}{3}(22-9C)\ep_i^2-(14-4C)\ep_i \eta_i +\fr{2}{3}\eta_i^2+\fr{1}{6}(13-3C)\xi_i
$$
\beq
r=16\ep_i\l 1-(\fr{2}{3}-2C)(2\ep_i-\eta_i)\r ~,
\la{r-ns-SRV}
\eeq
where we have still restricted the approximations up to the second order.
Applying these expressions, for the model (determining $\ep , \eta$ and $\xi $ as given in Eq. (\ref{ep-eta})),
we arrive at:
\beq
n_s=1\!-\!\l \!1+2\tan^2\fr{\al \phi_{\Te}^i}{2}\!\r \!(M_{Pl}\al )^2 \!+\!\l \fr{1}{6}\!+\!(1\!-\!\fr{1}{2}C)\tan^2\fr{\al \phi_{\Te}^i}{2}+
(\fr{1}{3}-\fr{1}{2}C)\tan^4\fr{\al \phi_{\Te}^i}{2}\r \!(M_{Pl}\al )^4 ~,
\la{n-model}
\eeq
and
\beq
r=8(M_{Pl}\al )^2\l 1-(\fr{1}{3}-C)(1+\tan^2\fr{\al \phi_{\Te}^i}{2})(M_{Pl}\al )^2\r \tan^2\fr{\al \phi_{\Te}^i}{2} ~.
\la{r-model}
\eeq
{}From Eq. (\ref{r-model}) we can express $\tan \fr{\al \phi_{\Te}^i}{2}$ in terms of $r$ and $M_{Pl}\al $. As will turn out, the latter's
value is small, so to a good approximation we find:
\beq
\tan^2 \fr{\al \phi_{\Te}^i}{2}\simeq \fr{r}{8(M_{Pl}\al )^2}\l \!1\!+\!(\fr{1}{3}\!-\!C)(\fr{r}{8}\!+\!(M_{Pl}\al )^2) \!+\!
\fr{1}{8}(\fr{1}{3}-C)^2(M_{Pl}\al )^4\r .
\la{aprox-phi-i}
\eeq
Plugging this into Eq. (\ref{n-model}) for the spectral index we get:
\beq
n_s-1=\!-\fr{r}{4}\!-\!(M_{Pl}\al )^2\!\!+\!\fr{1}{6}(M_{Pl}\al )^4-\fr{r^2}{64}(\fr{1}{3}-\fr{3}{2}C)\!+\!\fr{r}{8}(\fr{1}{3}+\fr{3}{2}C)(M_{Pl}\al )^2
\!+\!\fr{r^2}{128}(\fr{10}{9}-C(\fr{13}{3}-3C))(M_{Pl}\al )^2 .
\la{ns-rt}
\eeq
Using the recent value
 $n_s=0.968\pm 0.006$ from Planck \cite{Planck:2015xua}\footnote{The central value of $n_s$, is  larger, though the range (within $1\si $)
  is consistent with Planck's old result $n_s=0.9603\pm 0.0073$ \cite{Ade:2013zuv}. This required modification of our first version appeared
  before the new results.}${}^{,}$\footnote{Recent joint analysis of BICEP2/Keck and Planck \cite{Ade:2015tva} gave an upper bound $r\stackrel{<}{_\sim }0.12$, while
the likelihood curve for $r$ has a maximum for $r\simeq 0.05$. Note that  the value of $r$ reported before by BICEP2 collaboraton
 \cite{Ade:2014xna} was $r=0.2^{+0.07}_{-0.05}$ although, later on, Planck's intermediate results \cite{Adam:2014bub}
 warned about possible ordinary dust contribution instead of the light polarization effect really due to the gravitational waves.}
relation (\ref{ns-rt}) provides an upper bound for the
value of $M_{Pl}\al $:
\beq
M_{Pl}\al\stackrel{<}{_\sim }0.19 ~~~
({\rm obtained ~via ~ 2\si ~ variations ~of} ~n_s)~.
\la{Mplal-range}
\eeq
This will be used as orientation for further analysis and various predictions.

So far, we have performed calculations in a regime of small slow roll parameters, determining the value of $\phi_{\Te}^i$
via Eq. (\ref{aprox-phi-i}). As was mentioned, the value of  $\phi_{\Te}^e$ is determined from the condition $\ep_H=1$.
Near this point both $\ep $ and $\eta $ parameters turn out to be large
and instead of an expansion we need to perform numerical calculations.
 This will be relevant upon the calculation of the number of e-foldings $N_e^{\rm inf}$.

\begin{figure}
\begin{center}
\leavevmode
\leavevmode
\special{
psfile=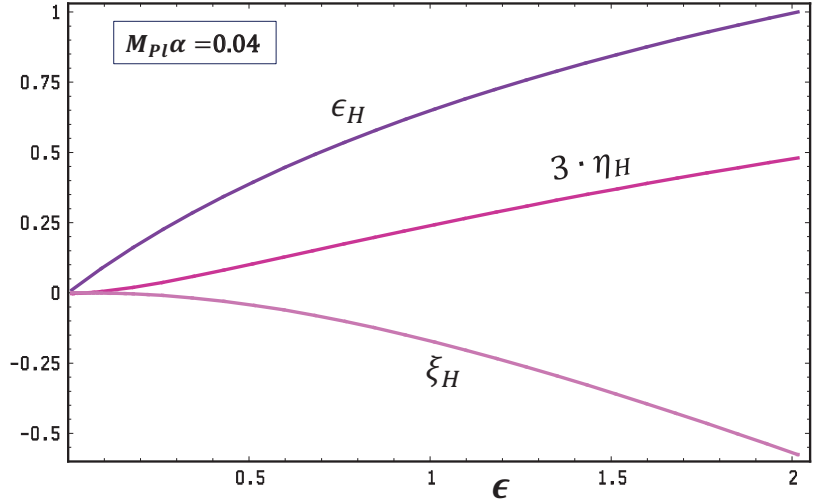 angle=0 voffset=-250
         hoffset=-260 hscale=60  vscale=60}  
\end{center}
\vs{6.1cm}
\caption{Dependence of $\ep_H, 3\eta_H$ and $\xi_H$ on the value of $\ep$, for $M_{Pl}\al =0.04$.}
\label{figHSR}
\end{figure}

\begin{figure}
\begin{center}
\leavevmode
\leavevmode
\special{
psfile=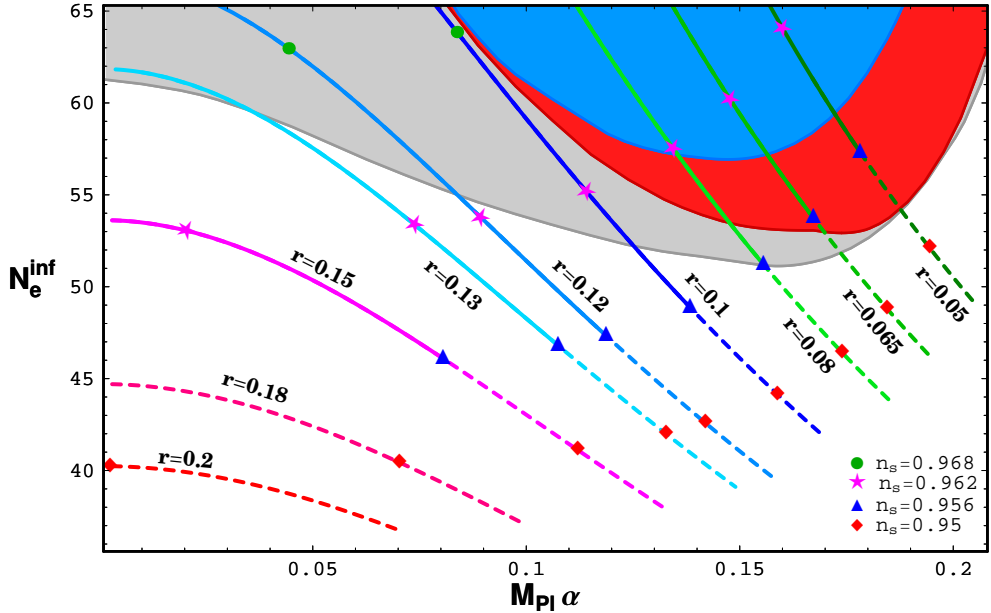 angle=0 voffset=-260 hoffset=-250 hscale=80  vscale=80

         }  
\end{center}
\vs{8.5cm}
\caption{Number of e-foldings.  Solid lines correspond to the values of $n_s$
which fit  with the current experimental data within $2\sigma $ error bars (with no restriction on $r$).
Shaded areas correspond to the marginalized joint $68\%$ CL regions, given recently in \cite{Planck:2015xua} for $(n_s, r)$ pairs,
mapped by us to  the $(M_{Pl}\al, N_e^{\rm inf})$ pairs for natural inflation. Gray background corresponds to the Planck TT$+$lowP,
while red and blue colors represent Planck TT$+$LowP$+$BKP and Planck TT$+$LowP$+$BKP$+$BAO respectively (see Ref. \cite{Planck:2015xua}
for an explanation of these combinations).}
\label{fig1}
\end{figure}

Since, within our model, via Eq. (\ref{ep-eta}) VSR parameters are related to each other as
\beq
\eta=\ep -\fr{1}{2}(M_{Pl}\al )^2~,~~~~\xi =-2(M_{Pl}\al )^2\ep ,
\la{rel-ep-eta-xi}
\eeq
the three equation in (\ref{exact-rel-HSR-VSR}) can be rewritten as
\beqr
\ep_H\l \fr{3-\eta_H}{3-\ep_H}\r^2&=&\ep \nonumber \\
\fr{3(\ep_H +\eta_H)-\eta_H^2-\xi_H}{3-\ep_H}&=& \ep -\fr{1}{2}(M_{Pl}\al )^2 \nonumber \\
3\fr{3-\eta_H}{(3-\ep_H)^2}\l 3\ep_H\eta_H+\xi_H(1-\eta_H)\r&=&-2(M_{Pl}\al )^2\ep~,
\la{model-rel-HSR-VSR}
\eeqr
where $\si_H$ has been dropped because  of it's  smallness. From the system of (\ref{model-rel-HSR-VSR}), for a fixed value
of $M_{Pl}\al $, the parameters $\ep_H, \eta_H$ and $\xi_H$ can be found in terms of the single parameter $\ep $.
The dependance of these parameters on the value of $\ep$, for   $M_{Pl}\al =0.04$ are shown in Fig. \ref{figHSR} (for different values of $M_{Pl}\al $
shapes of the curves are similar).
We see that
$\ep_H=1$ is achieved when $\ep=\ep_e\approx 2$ and thus, the expansion with respect to $\ep, \eta $ within this stage of inflation is invalid.
On the other hand, the values of $\eta_H$ and $\xi_H$ remain relatively small.
From the relation $2\ep =(M_{Pl}\al)^2\tan^2\fr{\al \phi_{\Te}}{2}$ one derives:
\beq
d\phi_{\Te}=\fr{M_{Pl}\sqrt{2}}{\sqrt{\ep}(2\ep+(M_{Pl}\al)^2)}d\ep .
\la{dPhi-deps}
\eeq
Using this, the integral in (\ref{exact-Ninf}) can be rewritten as
\beq
N_e^{\rm inf}=\int_{\ep_e}^{\ep_i}\!\!\!\fr{1}{2\ep+(M_{Pl}\al)^2}\fr{d\ep}{\sqrt{\ep \ep_H}} ~.
\la{exact-Ninf-1}
\eeq
Having the numerical dependence $\ep_H=\ep_H(\ep)$ (depicted in Fig. \ref{figHSR}),  we can
evaluate the integral in (\ref{exact-Ninf-1}) and find $N_e^{\rm inf}$ for various values of $M_{Pl}\al $. The results are given in Fig. \ref{fig1}.
While BICEP2/Keck and Planck \cite{Ade:2015tva}  reported the bound $r\stackrel{<}{_\sim}0.12$,
upon generating the curves of  Fig. \ref{fig1} we also allowed larger values of $r$.
Curves in Fig. \ref{fig1} and also Table \ref{tab1} demonstrate that, within natural inflation, with values $n_s\stackrel{<}{_\sim }0.962$
(the previous Planck 2013 value) and $r\stackrel{>}{_\sim }0.1$ or $n_s\stackrel{<}{_\sim }0.953$ and $r\stackrel{>}{_\sim }0.05$
there is an upper bound on $N_e^{\rm inf}$:
\beq
N_e^{\rm inf}\stackrel{<}{_\sim } 55 .
\la{bound-Ne}
\eeq
Turned around this also implies that violating the bound (\ref{bound-Ne}), say $N_e^{\rm inf}\sim 60$, indicates larger $n_s$ and/or smaller $r$.
Present Planck 2015 data seems to favor this.
Bound (\ref{bound-Ne}) (if realized) would lead to another striking prediction and constraint.

As discussed in  Refs. \cite{Liddle:2003as}, \cite{Ade:2013zuv}, the $N_e^{\rm eff}$, guaranteeing causality of fluctuations, should satisfy:
\beq
N_e^{\rm inf}=62-\ln \fr{k}{a_0H_0}-\ln \fr{10^{16}{\rm GeV}}{{\cal V}_i^{1/4}}+\ln \fr{{\cal V}_i^{1/4}}{{\cal V}_e^{1/4}}-
\fr{4-3\ga }{3\ga }\ln \fr{{\cal V}_{e}^{1/4}}{\rho_{reh}^{1/4}}~,
\la{Ne}
\eeq
where for the scale $k$  we take $k=0.002\hspace{0.6mm}{\rm Mpc}^{-1}$,
while the present horizon scale is $a_0H_0\approx 0.00033\hspace{0.6mm}{\rm Mpc}^{-1}$.
The factor $\ga $ accounts for the dynamics of the inflaton's oscillations  \cite{Turner:1983he}, \cite{NeferSenoguz:2008nn}
after  inflation, and can be for our model approximated
as $\ga \simeq 1-\fr{1}{16}\fr{{\cal V}_e}{{\cal V}_0}$ (will turn out to be a pretty good approximation).

To reconcile the first two entries ($62-\ln \fr{k}{a_0H_0}\approx 60.2$) of Eq. (\ref{Ne}) with the bound
of Eq. (\ref{bound-Ne}) (see also Fig. \ref{fig1}), the remaining entries of Eq.  (\ref{Ne}) should be significant enough to bring  $N_e^{\rm inf}$
down (at least) to $\approx 55$. The  $3^{\rm rd}$ and $4^{\rm th}$ entries on the r.h.s. of Eq.  (\ref{Ne}) can be calculated with help
of another observable - the amplitude of curvature perturbation $A_s$, which according to the Planck measurements \cite{Planck:2015xua},
\cite{Ade:2013zuv}, should satisfy
$A_s^{1/2}= 4.686 \tm 10^{-5}$ (this value corresponds to the $\La$CDM model).
 Generated by inflation, this parameter is given by:
\beq
A_s^{1/2}=\fr{1}{\sqrt{12}\pi }\left | \fr{{\cal V}^{3/2}}{M_{Pl}^3{\cal V}'}\right |_{\phi_{\Te}^i}\simeq
\fr{4\sqrt{6}}{3\pi }\fr{{\cal V}_0^{1/2}}{M_{Pl}^2}\fr{M_{Pl}\al}{r(1+8(M_{Pl}\al)^2/r)^{1/2}}~.
\la{dT-Ts}
\eeq
%
%
\begin{table}
\vs{-0.8cm}

 $$\begin{array}{|c|c|c|c|c|c|c|c|c|}

\hline
\vs{-0.4cm}
 &  &  &  &  &  &  & &  \\

\vs{-0.4cm}

r& M_{Pl}\al \hs{0.3mm}& \hs{0.3mm}n_s\hs{0.3mm}&\hs{-0.5mm} 10^4\!\tm \!\fr{dn_s}{d\ln k}\hs{-0.5mm} &\hs{0.3mm}N_e^{\rm inf}\hs{0.3mm}&
 \hs{0.5mm}\fr{{\cal V}_0^{1/4}}{10^{16}{\rm GeV}}\hs{0.5mm}&
\hs{0.5mm}\fr{{\cal V}_i^{1/4}}{10^{16}{\rm GeV}}\hs{0.5mm} &\hs{0.5mm}\fr{{\cal V}_e^{1/4}}{10^{16}{\rm GeV}}\hs{0.5mm}
&\hs{-0.5mm}\rho_{reh}^{1/4} ({\rm GeV})\hs{-0.5mm}\\

&  &  &  &  &  &  & &  \\

\hline
\hline

 &0.001  &0.962 &-7.1&53.62 &19.8  &2.01  &0.53  &3.14\!\cdot \!\!10^{4}    \\
\vs{-0.2cm}
 &0.04  &0.961 &-7.71&51.47 &3.2  &2.01  &0.54  &53.7  \\
\vs{-0.3cm}
  0.15&  &  &  &  &  &  &  & \\

 &0.055 &0.959&-8.25 &49.73 &2.77  &2.01  &0.54  &0.31   \\

 &0.065 &0.958& -8.71&48.39 &2.58  &2.01  &0.55  &0.006   \\
\hline
\hline
 &0.04 &0.967&-5.43 &61.25 &2.92  &1.92  &0.49  &2.77\!\cdot \!\!10^{14}    \\

  &0.06 &0.965& -6.06&57.92 &2.45  &1.92  &0.5  &1.39\!\cdot \!\!10^{10}    \\

 0.125  &0.08 &0.962 &-6.95&53.96 &2.2  &1.92  &0.52  & 1.14\!\cdot \!\!10^{5}  \\

 &0.1 &0.959& -8.09&49.79 &2.04  &1.92  &0.54  &0.5  \\

 &0.11 &0.957&-8.76 &47.72 &2  &1.92  &0.55  & 0.001 \\

\hline
\hline
 &0.097 &0.966& -5.53&60 &1.89  &1.81  &0.49  &9.2\!\cdot \!\!10^{12}    \\
  &0.1 &0.965&-5.68 &59.14 &1.87  &1.81  &0.5  &7.13\!\cdot \!\!10^{11}    \\

  0.1 &0.12 &0.961 &-6.79 &53.58 &1.79  &1.82  &0.52  & 5\!\cdot \!\!10^{4}  \\

    &0.135 &0.957 & -7.77&49.7 &1.74  &1.82  &0.54  &0.52 \\

    &0.143 &0.955 &-8.33&47.76 &1.72  &1.82  &0.55  &0.0017 \\

\hline
\hline
 &0.172 &0.958 &-4.53 &59.5 &1.35  &1.53  &0.47  &4.8\!\cdot \!\!10^{12}    \\
  &0.19 &0.952 &-5.36 &53.48 &1.34  &1.53  &0.49  &8.7\!\cdot \!\!10^{4}    \\
 \vs{-0.3cm}
  &0.194 &0.95 &-5.55 &50.93 &1.34  &1.53  &0.49  &44.8    \\
  \vs{-0.3cm}
  0.05   & & & & & &  & &  \\
 &0.2 &0.948 & -5.86&50.51 &1.33  &1.53  &0.5  & 13.6 \\

    &0.205 &0.946 &-6.12&49.12 &1.33  &1.53  &0.5  &0.225 \\

    &0.21 &0.944& -6.39 &47.78 &1.33  &1.53  &0.51  &0.0044 \\
    \hline
\end{array}$$
\caption{Numerical Results for  different values of $r$ and $M_{Pl}\al$. For all cases $A_s^{1/2}=4.686\tm 10^{-5}.$
 }
 \label{tab1}
\end{table}
%
%
%
In order to obtain the observed value of $A_s^{1/2}$, for typical  $r=0.12$ and  $M_{Pl}\al\sim 0.1$ we need to have
${\cal V}_0^{1/4}\sim 10^{-2}M_{Pl}$.
This, on the other hand, gives  ${\cal V}^{1/4}_i\sim  0.01M_{Pl}$ and
${\cal V}_e^{1/4}\sim 2\cdot 10^{-3}M_{Pl}$. Using these values in (\ref{Ne}) we see that the sum of the
 $3^{\rm rd}$ and $4^{\rm th}$ terms is$\approx 3.4$. Thus, the last term should be responsible for a proper reduction of $N_e^{\rm inf}$.
Namely, during the reheating process, the universe should expand by nearly $10$ (or even more) e-foldings.
This means that, for this case, the model should have a significant reheat history with
 $\rho_{reh}^{1/4}\sim 100$~GeV.\footnote{The reheating process can continue  even till  the epoch of nucleosynthesis.
 In this case one should have $\rho_{reh}^{1/4}\sim {\rm few}\tm 10^{-3}$~GeV.}
 Within the scenario of natural inflation, this has not been appreciated
 before.\footnote{See however some recent analysis in Ref. \cite{Munoz:2014eqa}.}
 For lower  $r$ and appropriate values of $M_{Pl}\al$  (and $n_s$)
  the reheating temperature can be big. The concise numerical results (compared to the rough evaluation below Eq. (\ref{dT-Ts})) are given in Table \ref{tab1}, where we considered
 cases with $\rho_{reh}^{1/4}$ not smaller than  $10^{-3}$~GeV, and $N_e^{\rm inf}\leq 62$.
  The values of the spectral index running $\fr{dn_s}{d\ln k}=16\ep_i\eta_i-24\ep_i^2-2\xi_i$ are also presented.
 The first three row-blocks correspond to the
 values of $n_s$ within $2\si $ ranges of the current experimental data.
 The first three cases of the bottom block correspond to the  $n_s$ within $3\si $ range, while the last three lines of this block
 have lower values of $n_s$ (beyond the $3\si $ deviation).
 Since the issue for the value of $r$ is not fully settled yet, we have included moderately large values of $r$($\leq 0.15$).
 At the bottom block of the table we gave
 results for $r=0.05$, which corresponds to the peak  of the $r$'s likelihood curve
 presented by the joint analysis of BICEP2/Keck and Planck \cite{Ade:2015tva}.
 Note that the results presented here are consistent with the analysis for natural inflation  carried out before
 \cite{Freese:2014nla} (see $2^{\rm nd}$ citation of this Ref.).

 Below we will show that within our scenario of natural inflation, a low reheat temperature is realized naturally.

\section{Natural inflation from 5D SUGRA}
\la{NATinf5D}

In order to address  the details of inflaton decay, related to the reheat temperature, we need to specify the underlying theory
natural inflation emerged from. A very good candidate is a higher dimensional construction \cite{ArkaniHamed:2003wu}.
Here we present a 5D conformal SUGRA realization \cite{Paccetti:2005zm} of this idea, using the off-shell superfield formulation developed in Refs.
\cite{Paccetti:2004ri}, \cite{ef4D-sugra}.\footnote{For the component formalism of  5D conformal SUGRA see the pioneering work
 by Fujita, Kugo and Ohashi \cite{Fujita:2001kv}.  Note also, that the component off shell 5D SUGRA formulation, discussed by Zucker \cite{zucker},  was used in many phenomenologically oriented papers.}

Lagrangian couplings, for the bulk ${\bf H}=(H, ~H^c)$ hypermultiplets', components are:
\beq
e_{(4)}^{-1}{\cal L}(H)=\hs{-0.2cm}\int d^4\te
(T+T^{\dagger })\l
H^{\dagger }H+H^{c\dagger }H^c \r +\hs{-0.2cm}
\int d^2\te \l 2H^c\pl_yH+g_1\Si_1(e^{{\rm i}\hat{\te }_1}H^2-
e^{-{\rm i}\hat{\te }_1}H^{c2})\r +{\rm h.c.}
\la{hyper-L}
\eeq
where the odd fields $V_i$ are set to zero.\footnote{The bulk hypermultiplet action of Eq. (\ref{hyper-L}), derived from 5D off shell SUGRA construction
\cite{Paccetti:2004ri}, including coupling with a radion superfield $T$, in a rigid SUSY limit coincides with the one given in Ref. \cite{Marti:2001iw}.}
$\Si_1$ is the $Z_2$ even $5^{\rm th}$ component of the 5D $U(1)$ vector supermultiplet.
With  the
parity assignments
\beq
Z_2:~~~H\to H~,~~~~~H^c\to -H^c~,
\la{1Hpar}
\eeq
 the KK decomposition for $H $ and $H^c$ superfields is given by
\beq
H=\fr{1}{2\sq{\pi R}}H^{(0)}+
\fr{1}{\sq{2\pi R}}\sum_{n=1}^{+\infty}H^{(n)}\cos \fr{ny}{R}~,~~~
~~
H^c=\fr{1}{\sq{2\pi R}}
\sum_{n=1}^{+\infty}\ov{H}^{\hs{0.6mm}(n)}\sin \fr{ny}{R}~.
\la{decKKphi}
\eeq
With these decompositions, and steps given in Appendix \ref{KK}, we can calculate the mass spectrum of KK states, their couplings to the inflaton and
with these,  the one loop order inflation potential (dropping higher winding modes)
having the form of (\ref{pot}) with
\beq
\al=\pi g_4R ,~~~~~~{\cal V}_0=\fr{3}{16\pi^6R^4}{\cal B} ~~~{\rm and}~~~{\cal B}=1-\cos(\pi R|F_T|)~.
\la{pot-5D-params}
\eeq
The 4D inflaton field $\phi_{\Te}$ is related to the 5D $U(1)$ gauge field $A_5^1$ as:
\beq
\phi_{\Te}=\sq{2\pi R}A_5^1~.
\la{rel-infl-A5}
\eeq

Since the model is well defined, we also can write down the inflaton coupling with the components of $H$. The latter, having a
coupling with the SM fields, would insure the inflaton decay and the reheating of the Universe.
In our setup, we assume that all MSSM matter and scalar superfields are introduced at the $y=0$ brane. Since $H$ is even under orbifold
parity and a singlet under all SM gauge symmetries, it can couple to the MSSM states through the following brane superpotential
couplings
\beq
{\cal L}_{H-br}=\sqrt{2\pi R}\int d^2\te dy\de(y)\lam lh_uH+{\rm h.c.}
\la{br-H}
\eeq
where $l$ and $h_u$ are  4D $N=1$ SUSY superfields corresponding to lepton doublets and up type higgs doublet superfields respectively.
In Eq. (\ref{br-H}), without loss of generality, only one lepton doublet (out of three lepton families) is taken to couple with the $H$,
$$
{\cal L}_{H-br}\supset -\lam \!\!\l (\fr{1}{\sqrt{2}}\psi_H^{(0)}+\sum_{n=1}^{+\infty}\psi_H^{(n)})(lh_u+\tl{l}\tl{h}_u)+
(\fr{1}{\sqrt{2}}H^{(0)}+\sum_{n=1}^{+\infty}H^{(n)})l\tl{h}_u \right.
$$
\beq
-\left.(\fr{1}{\sqrt{2}}F_H^{(0)}+\sum_{n=1}^{+\infty}F_H^{(n)})\tl{l}h_u\r +{\rm h.c.}
\la{br-H-1}
\eeq
where $l$ now denotes the fermionic lepton doublet and $h_u$  an up-type higgs doublet. States $\tl{l}$ and $\tl{h}_u$ stand for their
superpartners respectively. $H^{(n)}$ and $\psi_H^{(n)}$ in Eq.  (\ref{br-H-1}) indicate scalar and fermionic components of the
superfield $H$.\footnote{In Eq. (\ref{br-H-1}) we have omitted $HF_lh_u$ and $H\tl{l}F_{h_u}$ type terms, which because of the smallness of
the $\mu $ term($\sim $few$\tm $TeV) and suppressed lepton Yukawa couplings($\stackrel{<}{_\sim }10^{-2}$) can be safely ignored
in the inflaton decay proccess.}

Upon eliminating all $F$-terms and heavy fermionic and scalar states (in the $H$ and $H^c$ superfields), we can derive effective  operators
containing the inflaton linearly.
As it will turn out within  the model considered (see discussion in Appendix \ref{brane-K}), the $\tl l$ states are heavier than the inflaton and operators
containing  $\tl l$ are irrelevant for the inflaton decay. Thus, the  effective operators, needed to be considered, are
\beq
\phi_{\Te} \l C_0(lh_u)^2 +C_1(l\tl{h}_u)^2 +{\rm h.c.}\r
+C_2\phi_{\Te}(l\tl{h}_u)(\bar l\bar{\tl{h}}_u) .
\la{L-eff}
\eeq
These terms should be responsible for the inflaton decay.
Derivation and form of the $C$-coefficients are given in Appendix \ref{KK}.

\subsection{Inflaton Decay and Reheating}
\la{decay}

\begin{figure}[t]
\begin{center}
\hs{-1cm}
\resizebox{0.34\textwidth}{!}{
 \hs{0.5cm} \vs{0.5cm}\includegraphics{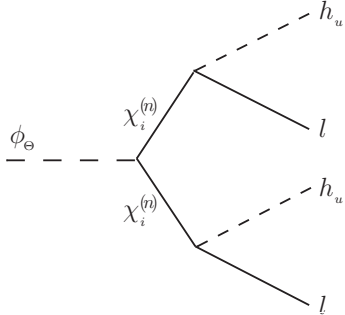}
}
\vs{0.2cm}
\caption{Diagram responsible for the  inflaton's dominant decay.}
\label{fig-decay}       
\end{center}
\end{figure}

As was mentioned above and
 shown in Appendix \ref{brane-K}, the slepton states $\tl{l}$ have masses $\fr{1}{2}|F_T|\sim 1/(2R)$ and thus are heavier than the inflaton.
Indeed, the latter's mass, obtained from the potential, is:
\beq
M_{\phi_{\Te}}=\fr{g_4\sqrt{3}(1-\cos(\pi R|F_T|))^{1/2}}{4\pi^2R} \ll \fr{1}{R}~.
\la{massInfl}
\eeq
($g_4\ll 1$ for successful inflation).
Thus the inflaton decay in channels containing $\tl l$ is kinematically
forbidden. Anticipating, we note that  the preheating process by inflaton decay in heavy states
is excluded within our scenario with parameters we consider (this is shown  at the end of this subsection).
Thus, the reheating proceeds by perturbative 4-body decay of the inflaton.

Among operators generated via exchange of heavy fermionic $\chi_i^{(n)}$
and scalar $S_i^{(n)}$ states, only those given in Eq. (\ref{L-eff}) are relevant.
For calculating the decay widths (in a pretty good approximation) it is enough to have the
form of the $C_i$ coefficients.

As shown in Appendix  \ref{KK}, within our model $C_2=0$ and the corresponding operator does not play any role. Moreover, according to Eqs. (\ref{C0-form}) and
 (\ref{C2-5-forms}) we have $C_0\sim R^2$ (due to a $1/m_{\psi_H}^2$) and $C_1\sim R^3$ (with $|F_T|\sim 1/R$, dictated from the inflation). Thus, we get an estimate for the
 following  branching ratio
 \beq
 \fr{\Ga (\phi_{\Te}\to ll\tl h_u\tl h_u)}{\Ga (\phi_{\Te}\to llh_uh_u)}\sim \fr{|C_1|^2M_{\phi_{\Te}}^7}{|C_0|^2M_{\phi_{\Te}}^5}
 \sim (RM_{\phi_{\Te}})^2\sim \fr{3g_4^2}{16\pi^4}\ll 1~.
 \la{branching}
 \eeq
 This means that the inflaton decay is mainly due to the $C_0$ operator [see Eqs. (\ref{L-eff} and (\ref{C0-form}), with gauge coupling $g_4$
 and Yukawa coupling $\lam$], i.e. in the channel $\phi_{\Te}\to llh_uh_u$ (the diagram in Fig. \ref{fig-decay}.
 Remember: $l$ denotes the SM lepton doublet and $h_u$ the scalar up type higgs doublet).
 For simplicity we assume that the state $h_u$ includes the light SM higgs doublet $h$ with weight nearly equal to one, i.e.
  $h_u\stackrel{\supset}{_\sim } h$.

For the decay width we get:\footnote{For 4-body phase space we have used an expression of \cite{Cappiello:2011qc}
derived for the $K\to \pi \pi ee$ decay, setting
$m_{\pi }, m_e\to 0$ and replacing $m_K\to M_{\phi_{\Te}}.$}

\beq
\Ga(\phi_{\Te})\simeq \Ga (\phi_{\Te}\to llh_uh_u)
=\fr{9}{9\cdot 2^7(2\pi)^5} |C_0|^2M_{\phi_{\Te}}^5 ~.
\la{infl-width}
\eeq
The factor $9$ in the numerator accounts for the multiplicity of final states. (The final $llh_uh_u$ channel
includes three combinations $e^-e^-h^+h^+$, $\nu \nu h^0h^0$, $e^-\nu h^+h^0$ and for each pair of identical final states
a factor $2$ should be included.)
The denominator factors in (\ref{infl-width}) come from the phase space integration.
Using the form of $C_0$, given by Eq. (\ref{C0-form}), in expression (\ref{infl-width}), we
get:
\beq
\Ga(\phi_{\Te})\simeq \fr{g_4^2|\lam |^4}{2^{18}\pi }(RM_{\phi_{\Te}})^4M_{\phi_{\Te}}~.
\la{infl-width-1}
\eeq

Expressing $\rho_{reh}=\fr{\pi^2}{30}g_*T_r^4$ through the reheat temperature \cite{Kofman:1997yn}
\beq
T_r=\l \fr{90}{\pi^2g_*}\r^{1/4}\!\!\sqrt{M_{Pl}\Ga(\phi_{\Te})}
\la{r-reh1}
\eeq
($g_*$ is the number of relativistic degrees at temperature $T_r$)
and using expressions (\ref{infl-width-1}) and (\ref{massInfl}), we get
\beq
\rho_{reh}^{1/4}= 1.316\l M_{Pl}\Ga(\phi_{\Te})\r^{1/2}=
5.85 \cdot 10^{-7}M_{Pl}\fr{g_4^{7/2}|\lam |^2}{(RM_{Pl})^{1/2}}(1-\cos(\pi R|F_T|))^{5/4} .
\la{r-reh2}
\eeq
From this, with $RM_{Pl}\sim 10$, $R|F_T|\sim 1$ and $g_4\sim 1.5\cdot 10^{-3}$ we obtain $\rho_{reh}^{1/4}\sim |\lam |^2\tm 100$~GeV.

%
%
%
\begin{table}
\vs{-0.8cm}

 $$\begin{array}{|c|c|c|c|c|c|c|}

\hline
\vs{-0.4cm}
& &  &  &  &  &     \\

\vs{-0.4cm}

(r,~ n_s)& M_{Pl}\al &  \hs{0.5mm}\fr{{\cal V}_0^{1/4}}{10^{16}{\rm GeV}}\hs{0.5mm} & \hs{0.3mm}R|F_T|\hs{0.3mm}&\hs{0.3mm}RM_{Pl}\hs{0.3mm}
& \hs{0.5mm}g_4\hs{0.5mm}&
\hs{0.5mm}\fr{\rho_{reh}^{1/4}}{|\lam |^{2}} ({\rm GeV})\hs{0.5mm}
\\

& & &  &  &  &    \\

\hline
\hline
\vs{-0.3cm}
& &  &0.75 &10.1 &1.26\!\cdot \!10^{-3}  & 60.5   \\
\vs{-0.3cm}
(0.15, 0.961)&  0.04& 3.2 &  &  &  &   \\

& &  &0.9 &10.5 &1.22\!\cdot \!10^{-3}   &62.6   \\
\hline
\hline
\vs{-0.3cm}
& &  &0.5 &10.2 &1.71\!\cdot \!10^{-3}  & 90.6   \\
\vs{-0.3cm}
(0.15, 0.959)&  0.055& 2.77 &  &  &  &   \\

& &  &0.9 &12.1 &1.45\!\cdot \!10^{-3}   &107   \\

\hline
\hline
\vs{-0.3cm}
& &  &0.5 &11 &1.88\!\cdot \!10^{-3}  & 122   \\
\vs{-0.3cm}
(0.15, 0.958)&  0.065& 2.58 &  &  &  &   \\

& &  &0.9 &13 &1.6\!\cdot \!10^{-3}   &145   \\
\hline
\hline
 \vs{-0.3cm}
& &  &0.5 &13.9 &2.29\!\cdot \!10^{-3}  & 216   \\
\vs{-0.3cm}
 (0.125, 0.959)& 0.1& 2.04 &  &  &  &   \\

& &  &0.9 &16.4 &1.94\!\cdot \!10^{-3}   &255   \\
    \hline
    \hline
 \vs{-0.3cm}
& &  &0.5 &16.3 &2.64\!\cdot \!10^{-3}  & 327   \\
\vs{-0.3cm}
(0.1, ~0.957)&  0.135& 1.74 &  &  &  &   \\

& &  &0.9 &19.3 &2.23\!\cdot \!10^{-3}   &386   \\
    \hline
    \hline

    \vs{-0.3cm}
& &  &0.5 &21 &2.91\!\cdot \!10^{-3}  & 409   \\
\vs{-0.3cm}
(0.05, ~0.95)&  0.194& 1.34 &  &  &  &   \\

& &  &0.9 &25 &2.45\!\cdot \!10^{-3}   &483   \\
    \hline
    \hline

 \vs{-0.3cm}
& &  &0.5 &21.3 &3\!\cdot \!10^{-3}  & 442   \\
\vs{-0.3cm}
(0.05, 0.948)&  0.2& 1.33 &  &  &  &   \\

& &  &0.9 &25.2 &2.53\!\cdot \!10^{-3}   &522   \\
 \hline
    \hline
 \vs{-0.3cm}
& &  &0.2 &14.1 &4.74\!\cdot \!10^{-3}  & 346   \\
\vs{-0.3cm}
(0.05, 0.944) & 0.21& 1.33 &  &  &  &   \\

& &  &0.9 &25.2 &2.65\!\cdot \!10^{-3}   &619   \\
    \hline
\end{array}$$
\caption{Values of $RM_{Pl}$, $g_4$ and $\rho_{reh}^{1/4}$ for different cases of successfull inflation.
 }
 \label{tab2}
\end{table}
%
%

Our 5D SUGRA construction allows more accurate estimates, because some of the parameters are related to each other.
For instance, from (\ref{pot-5D-params}) we have
\beq
R\simeq \fr{0.118}{{\cal V}_0^{1/4}}(1-\cos(\pi R|F_T|))^{1/4} ~,
\la{R-model}
\eeq
\beq
g_4=\fr{\al }{\pi R}~.
\la{g4-model}
\eeq
From (\ref{R-model}) we see that in order to have $RM_{Pl}\stackrel{>}{_\sim }10$ we need
${\cal V}_0^{1/4}\stackrel{<}{_\sim }3.4\cdot 10^{16}$~GeV.\footnote{For adequate
suppression of undesirable non local operators the large volume $R\stackrel{>}{_\sim }10/M_{Pl}$ is needed \cite{ArkaniHamed:2003wu}.}
The latter value suites well with most of the values of ${\cal V}_0^{1/4}$ given in Table \ref{tab1} (calculated from the inflation potential).
At the same time, we see from (\ref{R-model}) that $|F_T|$ can not be suppressed and should be $|F_T|\sim 1/R$.
Using Eqs. (\ref{R-model}) and (\ref{g4-model}) in (\ref{r-reh2}), we obtain
\beq
\rho_{reh}^{1/4}=
5.45 \cdot 10^{-5}M_{Pl}(\al M_{Pl})^{7/2}\l \fr{{\cal V}_0^{1/4}}{M_{Pl}}\r^{\!\!4}\hs{-0.1cm}|\lam |^2(1-\cos(\pi R|F_T|))^{1/4} .
\la{r-reh3}
\eeq
This expression is useful to find the maximal value of $\rho_{reh}^{1/4}$.
 Using the pairs of $(\al M_{Pl}, {\cal V}_0)$ given in Table \ref{tab1}, from Eq. (\ref{r-reh3}) it turns out that
  $\rho_{reh}^{1/4}\stackrel{<}{_\sim } |\lam |^2\tm 619$~GeV. This is an upper bound on the reheating energy density
  obtained within our 5D SUGRA scenario.
In Table \ref{tab2} we give
the values of $RM_{Pl}$, $g_4$ and $\rho_{reh}^{1/4}$ for various cases. Input values
of ${\cal V}_0^{1/4}$ and $M_{Pl}\al $ were taken from Table \ref{tab1}, which correspond to successful inflation.
Also, we have selected the values of   $R|F_T|$ in such a way as to get $RM_{Pl}\stackrel{>}{_\sim }10$.
We see that within $2\si $ deviations of $n_s$ we have $\rho_{reh}^{1/4}\stackrel{<}{_\sim } |\lam |^2\tm 386$~GeV,
 corresponding to reheat temperatures $T_r\stackrel{<}{_\sim } |\lam |^2\tm 130$~GeV.
These values can be easily reconciled with those low values of $\rho_{reh}^{1/4}$, given in Table \ref{tab1},
by natural selection of the brane Yukawa coupling $\lam $ in a range $1/300 \stackrel{<}{_\sim } \lam \stackrel{<}{_\sim }  1$.

\vs{0.3cm}

{\bf Excluding Preheating}

\vs{0.3cm}

\hs{-0.6cm}Since in the presented 5D SUGRA scenario the inflaton has direct couplings with heavy KK states of $H$ and $H^c$ superfields,
we need to make sure that after inflation, during the inflaton oscillation there is no production of these heavy states  and
no reheat is anticipated by the preheating process. Below we show that indeed, within our model preheating does not take place.

 Starting from the fermionic KK states (which turned out to dominate in reheating), their masses are given by Eq. (\ref{chi-masses}).
with $\lan M^1\ran =0$ and shift of the inflaton  around the vacuum
\beq
g_1A_5^1= g_1\lan A_5^1\ran +g_1\hat A_5^1=\fr{1}{R}+g_4\hat{\phi }_{\Te}~
\la{VEV-Sigma-0}
\eeq
for fermion masses we get
\beq
m_{\chi_1}^{(n)}=\fr{1}{2}\left |\fr{2n+1}{R}+g_4\hat{\phi }_{\Te}\right |~,~~~~
m_{\chi_2}^{(n)}=\fr{1}{2}\left |\fr{2n-1}{R}-g_4\hat{\phi }_{\Te}\right |~.
\la{chi-masses1}
\eeq
The $\hat{\phi }_{\Te}$ is the quantum part oscillating around the potential's minimum (after the end of inflation) and finally relaxing to $\hat{\phi }_{\Te}=0$.
Our aim is to see if either of the masses in (\ref{chi-masses1})  become zero during inflaton damped  oscillation. As was shown in
Ref. \cite{Giudice:1999fb}, this is the criterion for the fermionic preheating.

The amplitude of $\hat{\phi }_{\Te}$  has a well defined value at the end of the inflation when slow roll breaks down, i.e. at the point $\ep=\ep_f\simeq 2$.
 With $ \ep  =\fr{1}{2}(M_{Pl}\al)^2\cot^2 (\pi Rg_4 \hat{\phi }_{\Te}/2)$, for times $t>t_f$ we have
 \beq
 |g_4 \hat{\phi }_{\Te}^f|=\fr{1}{R} \fr{2}{\pi}{\rm ArcTan}\l \fr{M_{Pl}\al}{\sq{2\ep_f}}\r \simeq
 \fr{1}{R} \fr{M_{Pl}\al}{\pi } ~.
 \la{phi-hat-max}
 \eeq
 On the other hand, from our Table 2 we have $M_{Pl}\al\stackrel{<}{_\sim }0.2$. Using this in (\ref{phi-hat-max}), we get
 \beq
 |g_4 \hat{\phi }_{\Te}^f| \stackrel{<}{_\sim } \fr{0.064}{R}.
 \la{phi-hat-max1}
 \eeq
 The kinetic energy of the oscillation is still at most comparable at the end of inflation and there is also damping.
 Thus, Eq. (\ref{phi-hat-max1}) is a good estimate for the maximal amplitude of $\hat{\phi }_{\Te}$.
 With this bound, we can see that the term $g_4 \hat{\phi }_{\Te}$ in Eq. (\ref{chi-masses1}) will not be able
 to nullify fermion masses during inflaton oscillations. This fact, as was shown in Ref. \cite{Giudice:1999fb},
 prevents KK fermion production and no fermionic preheating takes place.

Now we turn to the scalar KK states. With Eqs. (\ref{S-masses}), (\ref{VEV-Sigma-0}) and $\lan M^1\ran =0$  for the scalar masses
we get
\beq
(m_{1,2}^{(n)})^2\!=\fr{1}{4}\!\l \fr{2n+1}{R}\pm |F_T|+g_4\hat{\phi }_{\Te}\r^2  ,~~~~~~
(m_{3,4}^{(n)})^2\!=\fr{1}{4}\!\l \fr{2n-1}{R}\mp |F_T|-g_4\hat{\phi }_{\Te}\r^2 ~.
\la{S-masses1}
\eeq
Therefore, with Eq. (\ref{phi-hat-max1}) and the values of $|F_T|$ given in Table 2, we see that masses in (\ref{S-masses1}) never cross zero
and for the positively defined mass$^2$ of the $S_i$ states we have
\beq
(m_i^{(n)})^2\stackrel{>}{_\sim }\l \fr{0.036}{R}\r^2\gg 10^4\tm (M_{\phi_{\Te}})^2~,
\la{S-mas-rel}
\eeq
where the inflaton mass $M_{\phi_{\Te}}\sim  10^{13}$~GeV.
Therefore, all modes from the scalar KK tower are much heavier (by a factor $>100$) than the inflaton mass for any time during the inflaton's oscillation.
As was shown in Refs. \cite{Kofman:1997yn}, \cite{Dufaux:2006ee} for these conditions the amplification and/or production of the scalar modes never happens.
 This excludes preheating also via the scalar production.
Within our scenario this result is insured by the gauge symmetry, because the inflaton in the heavy KK states' masses
contributes in the combination $g_4\hat{\phi }_{\Te}$ [see e.g. Eqs. (\ref{chi-masses1}) and (\ref{S-masses1})].

Thus, we finally conclude that within our scenario reheating occurs by the perturbative inflaton four-body decays discussed at the
beginning of Sec. \ref{decay}

\section{Discussion and Concluding Remarks}
\la{conc}

In the effective action of our 5D conformal SUGRA model the $5$-th component ($\Si_1 $) of a $U(1)$ vector
supermultiplet couples to  a charged hypermultiplet $(H, H^c)$. This, due to a fixed compactification radius $R$
leads to the potential of natural inflation for the CP odd part of $\Si_1$, neglecting the suppressed higher
winding modes. We analysed this potential like in \cite{Freese:2014nla} putting emphasis on the potential
of inflation and the number of e-folds of perturbations leaving the horizon.
This we compared with the number of e-folds required by a causal connection between the
observed universe background fluctuations and by the size of observed curvature perturbations.
For a large tensor component $r\stackrel{>}{_\sim }0.1$ and $n_s\stackrel{<}{_\sim }0.962$ or
 $r\stackrel{>}{_\sim }0.05$ and $n_s\stackrel{<}{_\sim }0.953$ close to the lower bound of present Planck 2015 data
 a small $N_e^{\rm inf}\stackrel{<}{_\sim }55$ results. This requires a small reheating temperature.
We inspected the decay of the gauge inflaton to the light MSSM fields living on a brane.
These decays are mediated by the bulk hypermultiplet $H$.
The very same $H$ hypermultiplet, together with it's $SU(2)_R$ partner $H^c$, generates the inflation potential.
The $H$ is assumed to have superpotential Yukawa couplings to brane fields with a Yukawa strength $\lam \stackrel{<}{_\sim }1$.
Due to a very small gauge coupling and to a
 relatively light  inflaton this led to a
 suppressed decay width and reheat temperature $T_r\sim |\lam |^2\tm 100$~GeV.
Within the considered scenario the dominant 4-body decays of the inflaton are mediated by fermionic components $\psi_H$ (of $H$)
with $llhh$ final states (two lepton and two higgs doublets' components).
 Other channels are either kinematically forbidden due to heavy sleptons $\tl l$ gaining
large masses through the large $F_T$ term, a case of split SUSY, or are suppressed (due to the  small inflaton mass $M_{\phi_{\Te}}\ll 1/R$) by
an additional small factor $(RM_{\phi_{\Te}})^2$.
Therefore, a similar mechanism can be realized also for extranatural inflation \cite{ArkaniHamed:2003wu} without
supersymmetry with a bulk fermionic   $\psi_H$ generating the inflation potential and brane Yukawa coupling $\lam lh\psi_H$.
Within our model (as shown in Appendix \ref{KK}), due to specific bulk couplings and degeneracy, the lepton number
is conserved and neutrinos stay massless (and this remains true for extra natural inflation).
 The situation can be changed by introducing a brane Majorana mass term $\fr{1}{2}M_{br}HH$
and it is inviting to exploit such a possibility. Since this is not directly related to inflation, on one side,
and trying to keep the calculus simple on the other side, we have not pursued this possibility in this paper
and reserved it for future studies.

The model of \cite{Paccetti:2005zm}, reanalyzed here in more detail, is by no means complete.
A concrete mechanism for radion stabilization like in Ref. \cite{Correia:2006vf} has to be presented
and the breaking of 4D SUSY has to be worked out in more detail.
Here and in \cite{Paccetti:2005zm} we concentrated on the aspects that our model originates in a very straightforward
way from 5D conformal SUGRA -which can be also interpreted as a result of M-theory \cite{ef4D-sugra}
 - and that the inflaton is related to a gauge field. If the new BICEP2 data, advertising large tensorial
fluctuations, will turn out not to be mainly dust effects and
if in the future, the value of $r$ turns out to be unsuppressed (i.e. $r\stackrel{>}{_\sim }0.05$ or so)
and if $n_s$ will not remain very close to the presently
favored higher values, then some form of
 natural inflation derived  from 5D gauge inflation would be indeed a suitable
and attractive candidate for inflationary model building.\footnote{See however footnote \ref{wgc} mentioning the 'weak
 gravity conjecture' and Ref. \cite{delaFuente:2014aca} proposing a 5D gauge field model
 leading to natural inflation without  need of a tiny gauge coupling.}
 If further findings will indicate a really small value of $r$, then
as an alternative, the 'modulus'
inflation of\cite{Paccetti:2005zm}  should be pursued. This would mean that the inflaton is the real part of the
$\Si_1$ chiral supermultiplet scalar component.
Also, a more general two field inflation \cite{2a} from complex $\Si_1$ could get into focus again.

\subsubsection*{Acknowledgments}

We thank Arthur Hebecker and Valeria Pettorino for discussions.
Research of Z.T. is partially supported by Shota Rustaveli National Science Foundation (Contracts No. 31/89 and No. DI/12/6-200/13).

\appendix

\renewcommand{\theequation}{A.\arabic{equation}}\setcounter{equation}{0}

\section{KK spectrum and the inflaton effective couplings}
\la{KK}

First let us discuss the emergence of the non zero $F_T$ term of the $T$ radion superfield. This can be easily understood
by the effective 4D SUGRA description developed in \cite{ef4D-sugra}. The 4D supergravity action is given by \cite{Kugo:1982mr}

$$
{\cal L}_{D}^{(4D)}+{\cal L}_F^{(4D)}~~~~~~~~{\rm with}
$$
\beq
{\cal L}_{D}^{(4D)}=- 3\int \!\!\!d^4\theta \,e^{-{\cal K}/3}\phi^+\phi ,   ~~~~~~~~~~~~~
{\cal L}_F^{(4D)}=\int \!\!\!d^2\theta\phi^3\, W +\frac{1}{4}\int \!\!\!d^2\te f_{IJ}{{\cal W}}^{\alpha I}{{\cal W}}_{\alpha}^J+ {\rm h.c.}
\label{eq:eleven}
\eeq
where  ${\cal K}$ and $W(\Phi)$  are   the \emph{K\" ahler} potential and the superpotential respectively,
while $f_{IJ}(\Phi)$ is the gauge kinetic function. $\phi$ is the 4D compensator chiral superfield.
Being a 4D effective theory,  (\ref{eq:eleven}) would include zero modes of the 5D supermultiplets and the brane fields as well.
Therefore, for the bulk states the form of   (\ref{eq:eleven}) will be dictated by the $5D$ construction \cite{ef4D-sugra}. For instance,
 the 4D compensator  $\phi$ is related with
the 5D compensator as $ \phi=\sqrt{2\pi R}\ka_5^{-\frac{1}{3}}h^{\frac{2}{3}}$.
From (\ref{eq:eleven}) we find the  expressions for the F-terms:
\beq
F^I=-M_Pe^{{\cal K}/3}{\cal K}^{I{\bar J}}D_{\bar J}{\bar W}, ~~~~
F_{\phi}=M_P^2e^{{\cal K}/3}\l {\bar W}-\fr{1}{3}{\cal K}^{I{\bar J}}{\cal K}_ID_{\bar J}{\bar W}\r .
\la{SUGRA-F-terms}
\eeq
where $I$ runs over all scalars.
 By plugging Eq. (\ref{SUGRA-F-terms}) back in  to (\ref{eq:eleven}),
one derives the $F$-term scalar potential (by setting $\phi =M_P$ and going to the 4D Einstein-frame, rescaling  the metric $g_{\mu \nu}\to e^{{\cal K}/3}g_{\mu \nu}$):
\be
 V_F=M_P^4\,e^{\cal K}\left({\cal K}^{I{\bar J}}D_IWD_{\bar J}{\bar W} - 3|W|^2\right), ~~~ {\rm with }~~~ D_I\equiv\partial_I+{\cal K}_I~.
\label{eq:formula}
\ee
For the $T$ modulus (the radion) the \emph{K\" ahler} potential  is ${\cal K}=-3\ln (T+T^\dag )$. For the time being we take $W=$const. for the superpotential.
\footnote{As shown in  \cite{Paccetti:2005zm}, this system (at zero mode level and for the purpose of discussing SUSY breaking) is equivalent to 5D SUGRA
with gauged $U(1)_R$ and with suitable couplings of a linear supermultiplet.}
With these, it is easy to check  that we get a flat potential $V_F=0$ with $F_{\phi }=0$ and $F_T=M_PW^*$. Thus, we have fixed a non zero $F_T$ which plays
a crucial role for the generation of the inflaton potential.
This is enough for performing a calculation of the KK spectrum and the 1-loop inflaton potential. We will come back to the SUSY breaking
at the end, upon discussion of the superpartners' spectrum from the MSSM brane fields.

Any bulk state transforming non trivially under $SU(2)_R$ feels $F_T$ SUSY breaking.
This happens of course with the bulk hypers described by the terms in (\ref{hyper-L}).
With the parametrization
\beq
F_T=-|F_T|e^{i\al }~,~~~~{\rm with}~~~~\al={\rm Arg}(F_T)+\pi ~,
\la{FT-phase}
\eeq
setting the scalar component of $T$ to one, and making a phase redefinition of the scalar components $H, H^c$:
\beq
H\to e^{-i(\hat{\te }_1+\al)/2}H~,~~~~~ H^c\to e^{i(\hat{\te }_1-\al)/2}H^c ~,
\la{HHc-phase}
\eeq
the couplings in (\ref{hyper-L}) give the potential:
$$
V(H)=2\left | \pl_5H-\fr{g_1}{2}(M^1-\fr{{\rm i}\Te }{2\pi R})H^c
-\fr{1}{2}|F_T|H^{c*}\right |^2
+2\left | \pl_5 H^c -\fr{g_1}{2}(M^1-\fr{{\rm i}\Te }{2\pi R})H
+\fr{1}{2}|F_T|H^*\right |^2+
$$
\beq
\fr{1}{2}g_1M^1|F_T|(H^2-H^{c2})+\fr{1}{2}g_1M^1|F_T|(H^{*2}-H^{c*2})~.
\la{Vphi}
\eeq

With the decomposition of Eq. (\ref{decKKphi}) and
 integrating along the fifth dimension $\int_0^{2\pi R}dy {\cal L}^{(5)}$ we obtain
 $$
 V(H^{(n)})=\left |g_1\Si_1H^{(0)}-\fr{1}{2}|F_T|H^{(0)*}\right |^2+
 \sum_{n=1}^{+\infty}\left |\fr{n}{R}H^{(n)}\!+g_1\Si_1\ov{H}^{(n)}+ \fr{1}{2}|F_T|\ov{H}^{(n)*}\right |^2+
 $$
 \beq
 \sum_{n=1}^{+\infty}\left |\fr{n}{R}\ov H^{(n)}\!-g_1\Si_1H^{(n)}+ \fr{1}{2}|F_T|H^{(n)*}\right |^2\!+
 \fr{1}{4}g_1M^1|F_T|\l \sum_{n=0}^{+\infty}(H^{(n)})^2-\sum_{n=1}^{+\infty}(\ov H^{(n)})^2+{\rm h.c.}\r
 \la{VHn}
 \eeq

 $$
 H^{(0)}=\fr{1}{\sq{2}}(S^{(0)}_1+iS^{(0)}_2)
 $$
 \beq
 {\rm for}~~~n=1,\cdots ,+\infty ~,~~~~~~
 \left(\!\!
   \begin{array}{c}
     {\rm Re}H^{(n)} \\
     {\rm Im}H^{(n)} \\
    {\rm Re}\ov H^{(n)} \\
      {\rm Im}\ov H^{(n)} \\
   \end{array}
 \!\!\right)\!\!=\!\fr{1}{2\sq{2}}
 \left(\!
   \begin{array}{cccc}
     1 & 1 & \!\!\!-1 & \!\!\!-1 \\
     \!\!\!-1 & 1 & 1 & \!\!\!-1 \\
     1 & \!\!\!-1 & 1 & \!\!\!-1 \\
     1 & 1 & 1 & 1 \\
   \end{array}
 \!\right)
 \!\!\left(\!\!
   \begin{array}{c}
   S^{(n)}_1 \\
   S^{(n)}_2 \\
   S^{(n)}_3 \\
   S^{(n)}_4 \\
   \end{array}
 \!\!\right)~,
 \la{scalar-basis}
 \eeq
( where $S^{(n)}_i$ are real scalars) and the potential mass terms will get diagonal and canonical forms:
\beq
V(S^{(n)})=\fr{1}{2}(m_1^{(0)})^2(S^{(0)}_1)^2+\fr{1}{2}(m_2^{(0)})^2(S^{(0)}_2)^2+
\fr{1}{2}\sum_{i=1}^4\sum_{n=1}^{+\infty}(m_i^{(n)})^2(S^{(n)}_i)^2~,
\la{VS}
\eeq
with
$$
(m_1^{(0)})^2=\fr{1}{4}(g_1A_5^1+|F_T|)^2+\fr{1}{4}(g_1M^1)^2 ~,~~~(m_2^{(0)})^2=\fr{1}{4}(g_1A_5^1-|F_T|)^2+\fr{1}{4}(g_1M^1)^2 ~,
$$
$$
{\rm for}~~~n=1,\cdots ,+\infty ~~:~~~~(m_{1,2}^{(n)})^2\!=\fr{1}{4}\!\l \fr{2n}{R}+g_1A_5^1\pm |F_T|\r^2\!\!\!+\fr{1}{4}(g_1M^1)^2 ,
$$
\beq
\hs{6.5cm}~~~~~(m_{3,4}^{(n)})^2\!=\fr{1}{4}\!\l \fr{2n}{R}-g_1A_5^1\mp |F_T|\r^2\!\!\!+\fr{1}{4}(g_1M^1)^2 ~.
\la{S-masses}
\eeq

As for the spectrum of the fermionic components of the $H, H^c$ superfields, with the phase redefinition
\beq
\psi_H\to e^{-i\hat{\te}_1/2}\psi_H ~,~~~~~\psi_{H^c}\to e^{i\hat{\te}_1/2}\psi_{H^c} ~,~~
\la{psi-phase}
\eeq
from Eq. (\ref{hyper-L}) we get the couplings
\beq
{\cal L}_{\psi}^{(5)}\supset -2\psi_{H^c}\pl_y\psi_H-\fr{1}{2}g_1(M^1-iA_5^1)(\psi_H\psi_H-\psi_{H^c}\psi_{H^c})+{\rm h.c.}
\la{L-psi}
\eeq
With the mode expansion of Eq. (\ref{decKKphi}) and integration over the fifth dimension
$\int_{0}^{2\pi R}dy{\cal L}_{\psi}^{(5)}={\cal L}_{\psi}^{(4)}$, from (\ref{L-psi}) we get terms
\beq
{\cal L}_{\psi}^{(4)}\supset-\sum_{n=1}^{+\infty}\fr{n}{R}\psi_{\ov H}^{(n)}\psi_{H}^{(n)}-
\fr{1}{4}g_1(M^1-iA_5^1)\l \sum_{n=0}^{+\infty}\psi_{H}^{(n)}\psi_{H}^{(n)}-
\sum_{n=1}^{+\infty}\psi_{\ov H}^{(n)}\psi_{\ov H}^{(n)}\r +{\rm h.c.}
\la{L-psi-KK}
\eeq
Now, with the substitution
$$
\psi_{H}^{(0)}=e^{i\om_0}\chi_1^{(0)}~,~~~~~{\rm with}~~~~\om_0=-\fr{1}{2}{\rm Arg}(M^1-iA_5^1)~,~~
$$
$$
{\rm for}~~~n=1,\cdots ,+\infty :~~~\psi_{H}^{(n)}\!=\!\fr{1}{\sq{2}}\!\l \!e^{i\om_n}\chi_1^{(n)}\!+\!e^{i\ov{\om}_n}\chi_2^{(n)}\!\r ,~~~
\psi_{\ov H}^{(n)}\!=\!\fr{i}{\sq{2}}\!\l \!-e^{i\om_n}\chi_1^{(n)}\!+\!e^{i\ov{\om}_n}\chi_2^{(n)}\!\r ,
$$
\beq
~~~~~~~~~~~{\rm with}~~~~\om_n=\!-\fr{1}{2}{\rm Arg}\!\l \!g_1M^1\!-i(g_1A_5^1+\fr{2n}{R})\!\r ~,~~
\ov{\om}_n=\!-\fr{1}{2}{\rm Arg}\!\l \!g_1M^1\!-i(g_1A_5^1-\fr{2n}{R})\!\r ~,
\la{psi-basis}
\eeq
from Eq. (\ref{L-psi-KK}) we will get diagonal and canonically normalized mass terms:
\beq
{\cal L}_{\psi}^{(4)}\supset -\fr{1}{2}\sum_{n=0}^{+\infty} m_{\chi_1}^{(n)}\chi_1^{(n)}\chi_1^{(n)}
-\fr{1}{2}\sum_{n=1}^{+\infty} m_{\chi_2}^{(n)}\chi_2^{(n)}\chi_2^{(n)}+{\rm h.c.}
\la{diag-chi-masses}
\eeq
with
\beq
m_{\chi_1}^{(n)}=\fr{1}{2}\left | g_1M^1\!-i(g_1A_5^1+\fr{2n}{R})\right |~,~~~~
m_{\chi_2}^{(n)}=\fr{1}{2}\left | g_1M^1\!-i(g_1A_5^1-\fr{2n}{R})\right |~.
\la{chi-masses}
\eeq

With this spectrum, integrating out the corresponding KK states (including zero modes) leads to the 1-loop effective potential
\cite{Paccetti:2005zm}, \cite{Hofmann:2003ag}:
$$
{\cal V}^{\rm eff}(\phi_{\Te })=\fr{3}{16\pi^6R^4}
\sum_{k=1}^{\infty} \fr{1}{k^5}\l 1-\cos (\pi kR|F_T|)\r \cdot
\cos(\pi kg_4R\phi_{\Theta })\tm
$$
\beq
e^{-\pi kg_4R|\phi_M|}\l 1+\pi kg_4R|\phi_M| +\fr{1}{3}(\pi kg_4R|\phi_M|)^2 \r ~,
\la{Vinfl}
\eeq
written in terms of canonically normalized 4D scalar
fields $\phi_{\Te }=\sq{2 \pi R}A_5^1$, $\phi_{M}=\sq{2\pi R}M^1$
and dimensionless 4D gauge coupling $g_4=g_1/\sq{2\pi R}$.
In (\ref{Vinfl}) summation is performed with $k$ winding modes. The dominant contribution comes
from $k=1$ \cite{Kohri:2014rja}. With this leading term,
 the minimum of the potential is achieved for $g_4R\lan \phi_{\Te }\ran =g_1R\lan A_5^1\ran =1$ and
 $\lan \phi_M\ran =0$. Further, we assume that the modulus $\phi_M$  (i.e. $M^1$) is sitting in its minimum and study only the motion of
 $\phi_{\Te }$'s quantum part as the inflaton. We add to the potential (\ref{Vinfl}) a constant term in such a way as
 to set the ground state vacuum energy to be zero (usual fine tuning of 4D cosmological constant). With these, the inflaton potential
 (part with $k=1$) gets the form of Eq. (\ref{pot}) with the parametrization given in Eq. (\ref{pot-5D-params}).

Further, we work out the effective couplings of the inflaton with the MSSM states. For this purpose,
in couplings (\ref{VS}), (\ref{chi-masses}) (and in any relevant term)
we make the substitution
\beq
g_1A_5^1\to g_1\lan A_5^1\ran +g_1A_5^1=\fr{1}{R}+g_4\phi_{\Te}~
\la{VEV-Sigma1}
\eeq
and put $\lan M^1\ran =0$. With this, we obtain
 the linear couplings of the inflaton with the heavy $S_i$  states:
$$
{\cal L}_{\phi_{\Te}SS}=-\fr{1}{2}g_4\phi_{\Te}\l \sum_{n=0}^{+\infty}[m_1^{(n)}(S^{(n)}_1)^2+m_2^{(n)}(S^{(n)}_2)^2]-
\sum_{n=1}^{+\infty}[m_3^{(n)}(S^{(n)}_3)^2+m_4^{(n)}(S^{(n)}_4)^2]\r ~,
$$
\beq
{\rm with}~~~~~~m_{1,2}^{(n)}=\fr{1}{2}\l \fr{2n+1}{R}\pm |F_T|\r ~,~~~~~m_{3,4}^{(n)}=\fr{1}{2}\l \fr{2n-1}{R}\mp |F_T|\r ~.
\la{phi-SS}
\eeq
At the same time, with (\ref{VEV-Sigma1}) from (\ref{psi-basis}) we have $\om_n=-\ov{\om}_n=\pi/4$, and Eq. (\ref{diag-chi-masses})
 gives inflaton  couplings with heavy $\chi_i$ states:
\beq
{\cal L}_{\phi_{\Te}\chi \chi}=\fr{1}{4}g_4\phi_{\Te}\l \sum_{n=0}^{+\infty}\chi_1^{(n)}\chi_1^{(n)} -\sum_{n=1}^{+\infty}\chi_2^{(n)}\chi_2^{(n)}\r
+{\rm h.c.}
\la{phi-chichi}
\eeq

Furthermore, we derive couplings of $S_i$ and $\chi_i$ states with the corresponding components of the brane superfields $l, h_u$.
As shown in Appendix \ref{brane-K}, the $\tl l$ states are heavy. Because of this, they will not be relevant for the inflaton
decay and we will omit any term containing the $\tl l$.
From the part of Eq. (\ref{br-H-1}) involving $\psi_H$ states we obtain
\beq
{\cal L}_{\chi lh_u}\supset -\fr{\lam e^{i(\pi-2\hat{\te}_1)/4}}{\sq{2}}\l \sum_{n=0}^{+\infty}\chi_1^{(n)}-i\sum_{n=1}^{+\infty}\chi_2^{(n)}\r
lh_u+{\rm h.c.}
\la{chi-lh}
\eeq
On the other hand,  making (\ref{HHc-phase}) phase  redefinitions, the part of Eq. (\ref{br-H-1}) involving $H$
gives:
\beq
{\cal L}_{Hlh_u} \supset -\fr{\lam e^{-i(\hat{\te}_1+\al)/2}}{\sq{2}}\l H^{(0)}+\sq{2}\sum_{n=1}^{+\infty}H^{(n)}\r l\tl{h}_u+{\rm h.c.}
\la{H-lh}
\eeq
From  Eq. (\ref{H-lh}) we get
 $S_il\tl{h}_u$ type couplings:
\beq
{\cal L}_{Slh_u} \supset \!-\fr{\lam}{2}e^{-i(\hat{\te}_1+\al)/2}\!\l \! S^{(0)}_1\!+iS^{(0)}_2+
e^{-i\pi/4}\!\sum_{n=1}^{+\infty}(S^{(n)}_1\!+iS^{(n)}_2\!-S^{(n)}_3\!-iS^{(n)}_4) \!\r \!l\tl{h}_u+{\rm h.c.}
\la{S-lh}
\eeq

Now, we integrate out the heavy $\chi_i$ and $S_i$ states, in order to obtain effective operators.
Starting with the integration of the fermionic modes,
at relatively low energies, we can ignore kinetic terms for the $\chi_i$ states. With this, via equations of motions
$\fr{\de {\cal L}}{\de {\chi_1}^{(n)}}=\fr{\de {\cal L}}{\de {\chi_2}^{(n)}}=0$, we can solve   ${\chi_{1,2}}^{(n)}$ and plug them back
into the Lagrangian. Doing so [using couplings of Eqs. (\ref{diag-chi-masses}), (\ref{phi-chichi}) and (\ref{chi-lh})] and keeping terms up to the
first power of $\phi_{\Te}$, we obtain:
\beq
\fr{i\lam^2}{4}e^{-i\hat{\te}_1}  \!\l   \sum_{n=0}^{+\infty}\l \fr{1}{m_{\chi_1}^{(n)}}-\fr{1}{m_{\chi_2}^{(n+1)}}\r  \!+
\fr{1}{2}g_4\phi_{\Te} \sum_{n=0}^{+\infty}\l  \fr{1}{(m_{\chi_1}^{(n)})^2}+\fr{1}{(m_{\chi_2}^{(n+1)})^2} \r \!\r \! (lh_u)^2+{\rm h.c.}
\la{L-d56-chi0}
\eeq
With $\lan M^1\ran =0, g_1\lan A_5^1\ran =1/R $, from (\ref{chi-masses}) we have $m_{\chi_1}^{(n)}=m_{\chi_2}^{(n+1)}=(2n+1)/(2R)$. Using this in (\ref{L-d56-chi0}),
we see that the first sum-term (coefficient in front of $d=5$ operator) cancels out, i.e. no $d=5$ lepton number violating operator emerges.
This is understandable, because the whole  theory has $U(1)$ gauge symmetry and the lepton number is a residual global symmetry
(with $\lan M^1\ran =0$) at $d=5$ level.
\footnote{Different result would emerge if we have had included brane $H^2$ coupling which explicitly violates the lepton number.
We do not consider such terms for the sake of simplicity.} Thus, from Eq. (\ref{L-d56-chi}) we obtain
\beq
{\cal L}^{d=6}_{(\chi)}=\fr{i\lam^2}{4}e^{-i\hat{\te}_1}
\l \sum_{n=0}^{+\infty} \fr{1}{(m_{\chi_1}^{(n)})^2}\r  g_4\phi_{\Te}(lh_u)^2+{\rm h.c.}
\la{L-d56-chi}
\eeq
where subscript $(\chi)$ indicates that this $d=6$  operator is obtained through the integration of the heavy  $\chi_i$ states.
The sum in (\ref{L-d56-chi}) is well convergent because $\sum_{n=0}^{+\infty} \fr{1}{(2n+1)^2}=\fr{\pi^2}{8}$.
It turns out that a $\phi_{\Te}(lh_u)^2$ type operator emerges only via integration of the $\chi_i$ states. Taking into account these, comparing
Eq. (\ref{L-d56-chi}) and (\ref{L-eff}) we have
\beq
C_0=\fr{1}{8}g_4i\lam^2e^{-i\hat{\te}_1}(\pi R)^2~.
\la{C0-form}
\eeq

Next, by integrating out heavy $S_i$ states, the $\phi_{\Te}(l\tl{h}_u)^2$ and $\phi_{\Te}(l\tl{h}_u)(\bar l\bar{\tl{h}}_u)$ type
dimension $7$ operators will be
$$
C_1\phi_{\Te}(l\tl{h}_u)^2 +{\rm h.c.}+ C_2\phi_{\Te}(l\tl{h}_u)(\bar l\bar{\tl{h}}_u) ~,
$$
$$
{\rm with}~~C_1=\fr{1}{8}g_4\lam^2e^{-i(\hat{\te}_1+\al)}\!\!\l \! -\!\!\sum_{n=0}^{\infty}\fr{e^{-i\fr{\pi}{2}(1-\de_{0n})}}{(m_1^{(n)})^3}
\!+\!\!\sum_{n=0}^{\infty}\fr{e^{-i\fr{\pi}{2}(1-\de_{0n})}}{(m_2^{(n)})^3}\!-
\! i\!\sum_{n=1}^{\infty}\fr{1}{(m_3^{(n)})^3}
\!+\! i\!\sum_{n=1}^{\infty}\fr{1}{(m_4^{(n)})^3}\!\r ~,
$$
\beq
C_2=\fr{1}{4}g_4|\lam |^2 \!\!\l \! \!-\!\sum_{n=0}^{\infty}\fr{1}{(m_1^{(n)})^3} \!-\!\sum_{n=0}^{\infty}\fr{1}{(m_2^{(n)})^3}
\!+\!\sum_{n=1}^{\infty}\fr{1}{(m_3^{(n)})^3} \!+\!\sum_{n=1}^{\infty}\fr{1}{(m_4^{(n)})^3}  \!\! \r .
\la{C2-5}
\eeq
Taking into account $m_3^{(n+1)}=m_2^{(n)}$ and  $m_4^{(n+1)}=m_1^{(n)}$, we see that the sums in $C_2$ precisely cancel out, while
$C_1$($\sim R^3$) remains non zero.
From the identity \cite{gandstein}
\beq
\sec^2\fr{\pi x}{2}=\fr{4}{\pi^2}\sum_{n=0}^{+\infty}\!\l \! \fr{1}{(2n+1+x)^2}  \!+\! \fr{1}{(2n+1-x)^2} \!\r ,
\la{sum-grands}
\eeq
taking first derivatives on both sides, we get
\beq
\sum_{n=0}^{+\infty}\!\l \! \fr{1}{(2n+1-x)^3}  - \fr{1}{(2n+1+x)^3} \!\r =\fr{\pi^3}{8} (1\!+\!\tan^2\fr{\pi x}{2})\!\tan \fr{\pi x}{2} ~.
\la{sum-for-C2}
\eeq
Using Eq. (\ref{sum-for-C2}), from (\ref{C2-5}) we finally  obtain:
$$
C_1=\fr{1}{4}g_4\lam^2e^{-i(\hat{\te}_1+\al)}\!\!\l \!  8\sq{2}e^{i\fr{\pi}{4}}R^4|F_T|\fr{3+(R|F_T|)^2}{(1-(R|F_T|)^2)^3}\!-
\! i(\pi R)^3\tan \fr{\pi R|F_T|}{2}(1+\tan^2 \fr{\pi R|F_T|}{2})\!\r ,
$$
\beq
C_2=0 ~.
\la{C2-5-forms}
\eeq
While the $C_2$ is precisely zero, the $C_1$ vanishes in the $F_T\to 0$ limit. However, with $|F_T|\sim 1/R$, we have $C_1\sim R^3$.

Remaining operators, as discussed in Sec. \ref{decay}, will not have any relevance for the inflaton decay and we will not
present them here.

\subsection{SUSY breaking on a brane}
\la{brane-K}

We assume that all MSSM states, that are matter $\{ f\}$, gauge $\{ V\}$ and higgs ${h_u, h_d}$ superfields, live on a 4D brane.
Matter superfields can be included in the   \emph{K\" ahler} potential as follows
\beq
{\cal K}=-\ln (T+T^\dag )^3-\ln \!\l \!1-\fr{2}{M_P^2}f^\dag e^{-V}f\!\r +{\cal K}(h_{u})+{\cal K}(h_{d})~,
\la{totalK}
\eeq
where ${\cal K}(h_{u,d})$ account for part of the higgs superfields and will be specified below.
With (\ref{totalK}), from  Eq. (\ref{eq:formula}), for squark and slepton masses we get
\beq
M_{\tl{f}}^2|\tl{f}|^2 ,~~~~{\rm with}~~~~M_{\tl f}^2=\fr{1}{4}M_P^2|W|^2=\fr{1}{4}|F_T|^2~.
\la{soft2}
\eeq
Due to the brane superpotential coupling of $l, h_u$ with $H$ state, there will be also a loop induced contribution to the soft mass$^2$, which we do not
display here.
Thus, with the large $F_T$-term all squark and sleptons are heavier than the inflaton field and they play no
role for the inflaton decay.
On the other hand we need to keep at least one Higgs doublet to be light. Since the SUSY breaking scale is very high, this can be achieved
only by price of fine tuning: assuming for instance that the light Higgs mainly resides in $h_u$, and selecting its
\emph{K\" ahler} potential as
\beq
{\cal K}(h_u)=\l 1+\al (T+T^\dag )^3\r \fr{2h_u^\dag e^{-V}h_u }{(1+8\al )M_P^2}~.
\la{Khu}
\eeq
Note, that with this selection, the kinetic term for $h_u$ is canonically normalized for arbitrary values of $\al$.
For the soft mass$^2$ of $h_u$ we obtain
\beq
M_{h_u}^2=\fr{27-(16\al +5)^2}{8(1+8\al)^2}|F_T|^2~.
\la{mass-hu}
\eeq
With the selection  $16\al+5=\sq{27}-{\cal O}(\fr{M_W^2}{F_T^2})$, we obtain
$M_{h_u}\sim {\cal O}(100 ~{\rm GeV})$ - the needed value.

As far as the gaugino masses are concerned, since the MSSM gauge supermultiplets are introduced on a brane
they will not have direct couplings neither with the $T$ modulus nor with the compensator.
By selecting, in Eq. (\ref{eq:eleven}), the gauge kinetic function $f_{IJ}=\de_{IJ}$, the corresponding gauginos will remain light.
By the same token, the higgsino mass - the $\mu $ parameter, can be around the TeV scale.
Therefore, the lightest neutralino can be a dark matter candidate.
 This  is the split SUSY scenario, which, as was shown \cite{ArkaniHamed:2004fb}, can have various remarkable phenomenological features
 and interesting implications.

%
%
%
%

\bibliographystyle{unsrt}

\newpage

\end{document}